\documentclass[twocolumn,aps,prd,epsf,nofootinbib]{revtex4-1}

\usepackage{color}
\usepackage{amsmath}
\usepackage{amssymb}
\usepackage{braket}
\usepackage[T1]{fontenc}
\usepackage{url}
\usepackage[english]{babel}
\usepackage{graphicx}
\usepackage{subfigure}
\usepackage{graphicx}
\usepackage{dcolumn}
\usepackage{bm}
\usepackage{float}
\usepackage{url}

\usepackage{bclogo}

\usepackage{tikz}
\usetikzlibrary{trees}
\usetikzlibrary{decorations.pathmorphing}
\usetikzlibrary{decorations.markings}

\usepackage{amsfonts} 

\usepackage{grffile}

\newcommand{\be}{\begin{equation}}
\newcommand{\ee}{\end{equation}}
\newcommand{\bea}{\begin{eqnarray}}
\newcommand{\eea}{\end{eqnarray}}

\newcommand{\bit}{\begin{itemize}}
\newcommand{\eit}{\end{itemize}}

\usepackage{verbatim}

\extrafloats{1000}

\graphicspath{{figs/}}

\begin{document}

\title{Lattice pure gauge compact QED in the Landau gauge: the photon propagator, the phase structure and the presence of Dirac strings}

\author{Lee C. Loveridge$^{1,2}$}
\email{loverilc@piercecollege.edu}
\author{Orlando Oliveira$^1$}
\email{orlando@uc.pt}
\author{Paulo J. Silva$^1$}
\email{psilva@uc.pt}

\affiliation{ \mbox{} \\ 
                 $^1$ CFisUC, Department of Physics, University of Coimbra, 3004-516 Coimbra, Portugal \\ \\
                 $^2$ Los Angeles Pierce College, 6201 Winnetka Ave.,  Woodland Hills CA 91371, USA}

\begin{abstract}
In this work we investigate the lattice Landau gauge photon propagator together with the average number of Dirac strings in the compact formulation 
of QED for the pure gauge version of the theory as a function of the coupling constant. Their $\beta$ dependence show that these two quantities
can be used to identify the confinement-deconfinement transition and that the nature of this transition is first order. Our results show that in the confined
phase the propagator is always finite, the theory has a mass gap and the number of Dirac strings present in the configuration is two orders of magnitude
larger than in the deconfined phase. Furthermore, in the deconfined phase where $ \beta \ge 1.0125$ the theory becomes massless, there are
essentially no Dirac strings and the photon propagator diverges when the limit $p \rightarrow 0^+$ is taken. Our results illustrate the importance
of the topological structures in the dynamics of the two phases.
\end{abstract}

\maketitle


\section{Introduction and Motivation}

The regularization of QED using a hypercubic lattice \cite{Wilson:1974sk} takes as fundamental fields the link variables
\begin{equation}
U_\mu(x) \equiv \exp\{ i \, e \, a\, A_\mu(x + a \, \hat{e}_\mu /2 ) \} \ ,
\label{Eq:Links}
\end{equation}
where $A_\mu$ is the bare continuum photon
field, $a$ is the lattice spacing, $e$ is the bare coupling constant and $\hat{e}_\mu$ is the unit vector along direction
$\mu$. The link variables $U_\mu(x)$ are defined on a compact manifold, they cover the unit circle centered around the origin of the complex plane.
On the other hand, the continuum photon field spans the real numbers. The phase diagram of the regularized compact formulation of QED 
has two phases that are distinguished by the value of the bare coupling constant or, equivalently, by $\beta = 1/e^2$. 

For low values of $\beta$, i.e in the strong coupling limit, the static potential between fermions fields grows linearly with the distance between the 
fermion sources and the theory is confining. For $\beta \gtrsim 1$, i.e. in the weak coupling limit, the static potential becomes
essentially constant at large distance separations, the theory is no longer confining,
and the results of the lattice formulation of QED approach those of the perturbative solution of the
continuum theory after taking the thermodynamic limit  
\cite{Banks:1977cc,Glimm:1977gz,Fradkin:1978th,Creutz:1979zg,Guth:1979gz,Lautrup:1980xr,Frohlich:1982gf,
 Berg:1983is,Kogut:1987cd,Coddington:1987yz,Nakamura:1991ww,Jersak:1996mj,Gubarev:2000eu,Arnold:2002jk,Panero:2005iu,Loveridge:2021qzs}. 
 
 The  confinement mechanism seems to be related to the topological structure of the Abelian gauge group $U(1)$ that, for the 3D case, is
 associated with the presence of Dirac monopoles. These classical configurations are absent in the deconfined phase but are observed in the confined
 phase
 \cite{Polyakov:1975rs,Polyakov:1976fu,Chernodub:2002gp,Bakker:2000kg,Greensite:2011zz}.
 Furthermore, in four dimensions, the confined phase has a mass gap \cite{Loveridge:2021qzs}.
 In the strong coupling limit compact QED has a mechanism that generates a photon mass gap\footnote{The photon propagator being finite in
the confined phase, means that its functional form should be of the type $Z(p^2)/( p^2 + M^2(p^2))$ and that the running photon mass
$M^2(p^2)$ does not vanish in the infrared region. 
In the main text when we do refer to a photon mass we have in mind the function $M(p^2)$. 
Recall that in the deconfined phase the propagator is compatible with a divergent behaviour that implies a vanishing $M^2(p^2)$ at zero momentum. 
It is in this sense that the deconfined phase is understood as being a massless theory.} and the theory is confining. 

The presence of a mass gap is also observed in the simulation of non-Abelian gauge theories.
Indeed, the generation of mass scales was observed in non-Abelian gauge theories as in QCD and in the pure Yang-Mills SU(2) gauge theory
\cite{Alkofer:2003jj,Cucchieri:2007md,Fischer:2008uz,Aguilar:2008xm,Boucaud:2008ky,Dudal:2008sp,Bogolubsky:2009dc,Dudal:2010tf,
Oliveira:2012eh,Duarte:2016iko,Cucchieri:2016jwg,Cyrol:2016tym,Dudal:2018cli,Huber:2018ned,Oliveira:2018lln,Gracey:2019xom,Dudal:2019gvn,
Huber:2020keu,Fischer:2020xnb,Falcao:2020vyr,Aguilar:2021lke,Gao:2021wun,Lo:2021qkw,Hayashi:2021nnj}. 
 
The understanding of the confinement mechanism for non-Abelian gauge theories and its possible connection with the generation of mass gaps
is still a fundamental open problem for Particle Physics. Hopefully,  the comparison of the compact QED formulation with the non-Abelian case will 
bring further insight into the confinement problem.

The dynamical properties of compact QED depend on the value of $\beta$. This dependence translates into deep changes in the 
properties of the QED Green functions, i.e. in the propagators and vertices of the theory. Indeed, 
the computation of the photon propagator in the Landau gauge for pure gauge lattice compact QED  at low $\beta$ and at high $\beta$
illustrates those differences \cite{Berg:1983is,Nakamura:1990dq,Nakamura:1991ww,Durr:2002jc,Loveridge:2021qzs}.
For the confined low $\beta$ phase, the photon propagator seems to be described by a Yukawa-type of propagator. 
The theory has a mass gap and the topological structures, measured by the average number of Dirac strings that are observed in the gauge configurations,
are orders of magnitude higher than in the large $\beta$ phase where confinement is not present. 
Furthermore,  in the deconfined high $\beta$ phase the  photon propagator seems to  approach the continuum like $1/p^2$ behaviour of a free field theory 
as we go towards higher volumes.
 
 In the current work we revisit the Landau gauge photon propagator computed using the lattice formulation of pure gauge compact QED in 
 four dimensions and study its $\beta$ dependence. The investigations reported here complement the recently published paper \cite{Loveridge:2021qzs}
 that illustrates how different the photon propagator is in the confined and deconfined phases. 
 
Our study shows that in the confined phase the photon propagator is finite over the full range of momenta, i.e. it has a mass gap, 
while in the deconfined phase it seems to diverge as $p \rightarrow 0$. For the deconfined phase the zero momentum propagator
increases with $\beta$.
The  average number of Dirac strings present in the gauge configurations is also phase dependent. The number of Dirac strings
is much larger in the confined phase,  in comparison  with the deconfined phase where the Dirac strings are almost non-existent. 
Moreover, the average number of Dirac strings is strongly correlated with the properties of the photon propagator and this correlation suggests
that the origin of the photon mass observed in the confined phase is due to these topological structures. 
Although we do not compute the photon mass as a function of the coupling constant, as this would require modelling the propagator,
in the confined phase the propagator is finite for all momenta a result that can be viewed as due to the generation of a finite non-vanishing 
running mass for the photon field. In the same sense, the divergence of the photon propagator at zero momentum in the deconfined phase can 
be read as an indication that the theory becomes massless.

Our results show that the transition between the two phases can be clearly identified by looking either at the photon propagator data or at
the average number of Dirac strings. From either of these two quantities it is possible to estimate the corresponding critical value of $\beta_c$  
where the phase transition occurs. In this sense, the analysis of the propagator and/or the number of Dirac strings can be used as order parameters 
for the confinement-deconfinement transition for pure gauge compact QED. 
Further, the photon propagator and the Dirac strings data show that the transition to the deconfined phase 
is of first order in good agreement with the literature \cite{Fradkin:1978th,Guth:1979gz,Lautrup:1980xr,Frohlich:1982gf,Kogut:1987cd,Arnold:2000hf}. 

The understanding of the phase diagram of compact QED is important \textit{per se}. Historically QED is the first quantum field theory to be studied
and has furnished a number of results and methods that were later generalized to other types of theories.
Also QED is a part of the Standard Model of Particle Physics and the interplay of electrodynamics with strong interactions is currently a subject of 
research. Furthermore, it is essential to have a good theoretical understanding of QED to perceive the Higgs sector of the Standard Model
and of Abelian Higgs models, that are also relevant for Condensed Matter systems. QED has a fundamental role in 
Condensed Matter Physics and Atomic Physics and $U(1)$ gauge theories are also being used to explore the possibility of performing 
realistic simulations with quantum computers.

The paper is organized as follows. In Sec. \ref{Sec:setup} we define the theoretical setup of our study, giving details on the gauge fixing and on the
computation of the photon propagator. In Sec \ref{SecProp} we report on the photon propagator results for various $\beta$ and how the propagator
evolves with $\beta$. 
In Sec. \ref{Sec:DiracStrings} the Dirac string content and distributions in the different phases of the theory is analyzed.
The correlation of these topological structures with the photon propagator data is also discussed there. 
Issues related to the performance of the sampling with the hybrid Monte Carlo method are also touched.
Finally, in Sec. \ref{Sec:Summary} we summarize and conclude.

\section{Pure Gauge Compact QED \label{Sec:setup}}

The simulations considered herein refer to the compact version of QED defined over an hypercubic
lattice and described by the Wilson action. In Euclidean space the Wilson action reads
\begin{equation}
   S_W (U) = \beta \sum_x \sum_{1 \leqslant \mu, \nu \leqslant 4} \bigg\{ 1 - \Re \, \left[ U_{\mu\nu} (x) \right]\bigg\} \ ,
   \label{Eq:accao}
\end{equation}
where the plaquette operator is given by
\begin{equation}
   U_{\mu\nu} (x) = U_\mu (x) \, U_\nu (x + a \, \hat{e}_\mu) \, U^\dagger_\mu (x + a \, \hat{e}_\nu) \,  U^\dagger_\nu (x) 
   \label{Eq:plaquette}
\end{equation}
that, in the continuum limit, is
\begin{equation}
   U_{\mu\nu} (x) = \exp\left\{ i \, e \, \oint_C A_\mu (z)  \, dz_\mu \right\}
\end{equation}
with $C$ being any closed curve that contains in its interior the point $x$ and whose points are infinitesimally close to $x$. 
On an hypercubic lattice whose lattice spacing is $a$, the link variables are related to the photon field $A_\mu$ by
Eq. (\ref{Eq:Links}).
It follows from the definition given in Eq. (\ref{Eq:plaquette}) that the exponential term is the change of the photon field 
around a plaquette centered at  $x + a \, ( \hat{e}_\mu + \hat{e}_\nu ) /2$. Indeed, writing for the plaquette
\begin{equation}
   U_{\mu\nu} (x) = \exp\left\{ i \, e \, a \, \Big( \, \Delta A_{\mu\nu} (x) \, \Big)  \right\} \ ,
\end{equation}
given that  $- \pi \leqslant e \, a\, A_\mu \leqslant \pi$ and that 
$- \pi \leqslant e \, a ~ \Delta A_{\mu\nu} (x) \leqslant \pi$, it follows that
\begin{eqnarray}
\Delta A_{\mu\nu} (x) & = &
 A_\mu \left( x + \frac{a}{2} \, \hat{e}_\mu \right) +
                    A_\nu \left( x + a \, \hat{e}_\mu + \frac{a}{2} \, \hat{e}_\nu \right) \nonumber \\
                    & & \quad -
                    A_\mu \left( x +  a \, \hat{e}_\nu + \frac{a}{2} \, \hat{e}_\mu  \right) -
                   A_\mu \left( x + \frac{a}{2} \, \hat{e}_\nu \right)   \nonumber \\
                   & & \qquad\quad + \, \frac{ 2 \, \pi \, m_{\mu\nu} (x)}{e \, a }
                   \label{EqDirac_string0}
\end{eqnarray}
where the integer field $m_{\mu\nu}(x)$ measures the number of Dirac strings that cross the plaquette associated with 
$U_{\mu\nu} (x)$. In a simulation, for a given gauge configuration, the  field $m_{\mu\nu}(x)$ can be 
computed by combining links and plaquettes as described in \cite{DeGrand:1980eq,Loveridge:2021qzs}. 
As was shown in \cite{Loveridge:2021qzs}, the confined and deconfined phases can be distinguished by looking at $m_{\mu\nu}(x)$ 
or to its average over the lattice.

In the simulations reported below, the links were sampled with the Wilson action (\ref{Eq:accao}),
relying on the hybrid Monte Carlo method (HMC)  \cite{Duane:1987de} that was implemented with the Chroma library that requires 
the QDP++ library \cite{Edwards:2004sx}. 
It is well known that the HMC has problems simulating compact QED \cite{Grosch:1985cz,Kerler:1995va},
as the various topological sectors are not properly sampled. 
The topological freezing of the HMC becomes more important close to the transition between the confined and 
deconfined region and also in the deconfined phase. 

In an effort to estimate the bias introduced by the sampling method, close and above the transition between the two phases, i.e. for $\beta \ge \beta_c$,
several simulations were performed using the same $\beta$ and
taking different starting points  to initiate the Monte Carlo. 
We also considered long runs for some of the $\beta$'s. 
Our results show that for $\beta \le \beta_c$ the different Markov chains result in the same photon propagator.
However, for $\beta \ge \beta_c$ the photon propagators differ between the different simulations. A correlation between the photon propagator and the 
average value of Dirac strings is observed; see the results of Sec \ref{Sec:HMC-DiracStrings} for details. 
For $\beta \ge \beta_c$ the propagators for the different simulations with the same $\beta$ value overlap in the UV region but show
different functional forms in the infrared region. We also also find that for $\beta \ge \beta_c$ and for the different simulations
with the same $\beta$, the photon propagator resulting from the configurations with the smaller number of Dirac strings is closer to the perturbative propagator.
If for the deconfined high $\beta$ phase the photon propagator obtained with the hybrid Monte Carlo method shows
a dependence on the initial configuration in the infrared region,
the overall analysis of the $\beta$ dependence is robust and does not change qualitatively
by taking either of the photon propagators computed for $\beta > \beta_c$. On the other hand, for the confined low $\beta$ phase our results suggest that the HMC is robust. For the deconfined high $\beta$ phase our simulations confirm the sampling problems of the hybrid Monte Carlo method previously identified in the literature.

After importance sampling, the links were rotated towards the Landau gauge as in \cite{Loveridge:2021qzs}.
On a first stage, we rely on the linear definition for the photon field given by
\begin{equation}
   e \, a \, A_\mu \left( x + \frac{a}{2} \hat{e}_\mu \right)  = \frac{U_\mu(x) - U^\dagger_\mu(x)}{2 \, i}
   \label{Eq:GaugeFieldLinear}
\end{equation}
and maximize the functional
\begin{equation}
   F[U; g] = \frac{1}{V \, D} \sum_{x,\mu} \, \Re \left[  \,  g(x) \, U_\mu (x) \, g^\dagger(x + a \, \hat{e}_\mu )\,  \right] 
   \label{Eq:Landau_linear}
\end{equation}   
over the gauge orbit. In Eq. (\ref{Eq:Landau_linear}) $g(x) \in U(1)$, $V$ is the total number of lattice points and $D$ the Euclidean spacetime dimension. 
In a second stage, the photon field is computed with the logarithmic definition
\begin{equation}
   e \, a \, A_\mu \left( x + \frac{a}{2} \hat{e}_\mu \right) = -i \, \ln \Big(  U_\mu (x) \Big) \ ,
   \label{Eq:photon_field}
\end{equation} 
that provides an exact definition, up to machine precision, and does not rely on the use of a small lattice spacing.
Then, the Landau gauge condition is achieved by maximizing, over the gauge orbits, the functional  \cite{Ilgenfritz:2010gu} 
\begin{equation}
   \widetilde{F}[U;g] = \frac{1 }{V \, D } \sum_{x ,\mu} \,  \bigg\{ 1 -  a^2 e^2 \left[ A^{(g)} _\mu \left( x + \frac{a}{2} \hat{e}_\mu \right) \right]^2 \bigg\}
   \label{Eq:Landau_Log}
\end{equation}
where the field $e \, a \, A^{(g)}$ is the photon field as given by Eq. (\ref{Eq:photon_field}) using the links $U_\mu(x)$ gauge transformed by 
$g(x)$.  The gauge fixing towards the Landau gauge is, in all cases, monitored computing the lattice version of $\partial \cdot A$
and, for each stage, the gauge fixing was stopped for an averaged value over the lattice of $|\partial \cdot A|^2 < 10^{-15}$.
Further details on the gauge fixing can be found in \cite{Loveridge:2021qzs}.

The optimizing functions that define the Landau gauge on the lattice have multiple maxima that lead to different Landau gauge configurations,
the Gribov copies. In all the simulations we ignored the various maxima and the Landau gauge fixing was performed starting the iterative 
process with $g(x) = 1$ and performing a single maximization for each gauge configuration obtained with the Monte Carlo sampling of the Wilson action.

From the link variables, the  momentum space photon field is computed with the defnition
\begin{equation}
  A_\mu (p) =  \sum_x \, e^{ -i p \cdot( x + \frac{a}{2} \hat{e}_\mu )}  A_\mu \left( x + \frac{a}{2} \hat{e}_\mu \right) \ ,
\end{equation}
using  the spacetime photon field given  in (\ref{Eq:photon_field}). The Landau gauge propagator is defined as
\begin{equation}
   \langle A_\mu (p_1) ~ A_\mu(p_2) \rangle = V \, \delta( p_1 + p_2 ) \, D_{\mu\nu}(p_1)
   \label{Eq:prop1}
\end{equation}
where $\langle \cdots \rangle$ stands for the average over the gauge fields. The analysis of the propagator is performed
assuming that the propagator has the same tensor structure as in the continuum
\begin{equation}
D_{\mu\nu}(p) = \left( \delta_{\mu\nu} - \frac{p_\mu p_\nu}{p^2} \right) \, D(\hat{p}) \ .
\label{Eq:photon_prop}
\end{equation}
The form factor $D(\hat{p})$ is a function of the tree level improved momenta
\begin{eqnarray}
  & &
    \hat{p} = \frac{2}{a} \sin \left( \frac{\pi}{L} \, n_\mu \right) \ , \nonumber \\
    & &  \quad\quad
     n_\mu = -\frac{L}{2}, \, -\frac{L}{2} + 1, \, \dots , \, 0, \, 1, \, \dots , \, \frac{L}{2}  - 1
\end{eqnarray}    
where $L$ is the number of lattice points in each side of the hypercubic lattice. The rationale for 
using $ \hat{p}$ instead of the naive lattice momenta $p =  2 \pi  n_\mu / a  L$ is to reduce the finite space
effects in the propagator \cite{Leinweber:1998uu,Catumba:2021hcx}, a procedure developed for asymptotically free field theories.  
The lattice data for the propagator shown here satisfies the conical and cylindrical momentum cuts  \cite{Leinweber:1998uu} for 
$a \, \hat{p} > \Lambda_{\text{IR}}$. These momentum selections  were set to further suppress the finite space effects and to produce 
a well defined curve from the lattice data. 
For $a \, \hat{p} \leqslant \Lambda_{\text{IR}}$ we follow \cite{Dudal:2018cli} and consider all the momenta accessed in the simulations.
In all cases we   use $\Lambda_{\text{IR}} = 0.4$.

Assuming that the propagator tensor structure is as given in Eq. (\ref{Eq:photon_prop}), then 
\begin{equation}
D(\hat{p}) = \left\{ 
\begin{array}{l @{\hspace{1cm}} l}
    \frac{1}{3} \sum_{\mu = 1}^4 D_{\mu\mu} (p) , & \hat{p} \ne 0  \ , \\ 
    & \\
    \frac{1}{4} \sum_{\mu = 1}^4 D_{\mu\mu} (p) , & \hat{p} = 0  \ .
   \end{array}
   \right.
\end{equation}

The propagators for the different $\beta$ values were computed on a $48^4$ hypercubic lattice using the last (in the Markov chain) 200 
gauge configurations. 
The statistical errors are evaluated with the bootstrap method for a confidence level of 67.5\%, except for the fits where we use 
Gaussian error propagation.

\begin{figure*}[t] 
   \centering
   \includegraphics[width=3.2in]{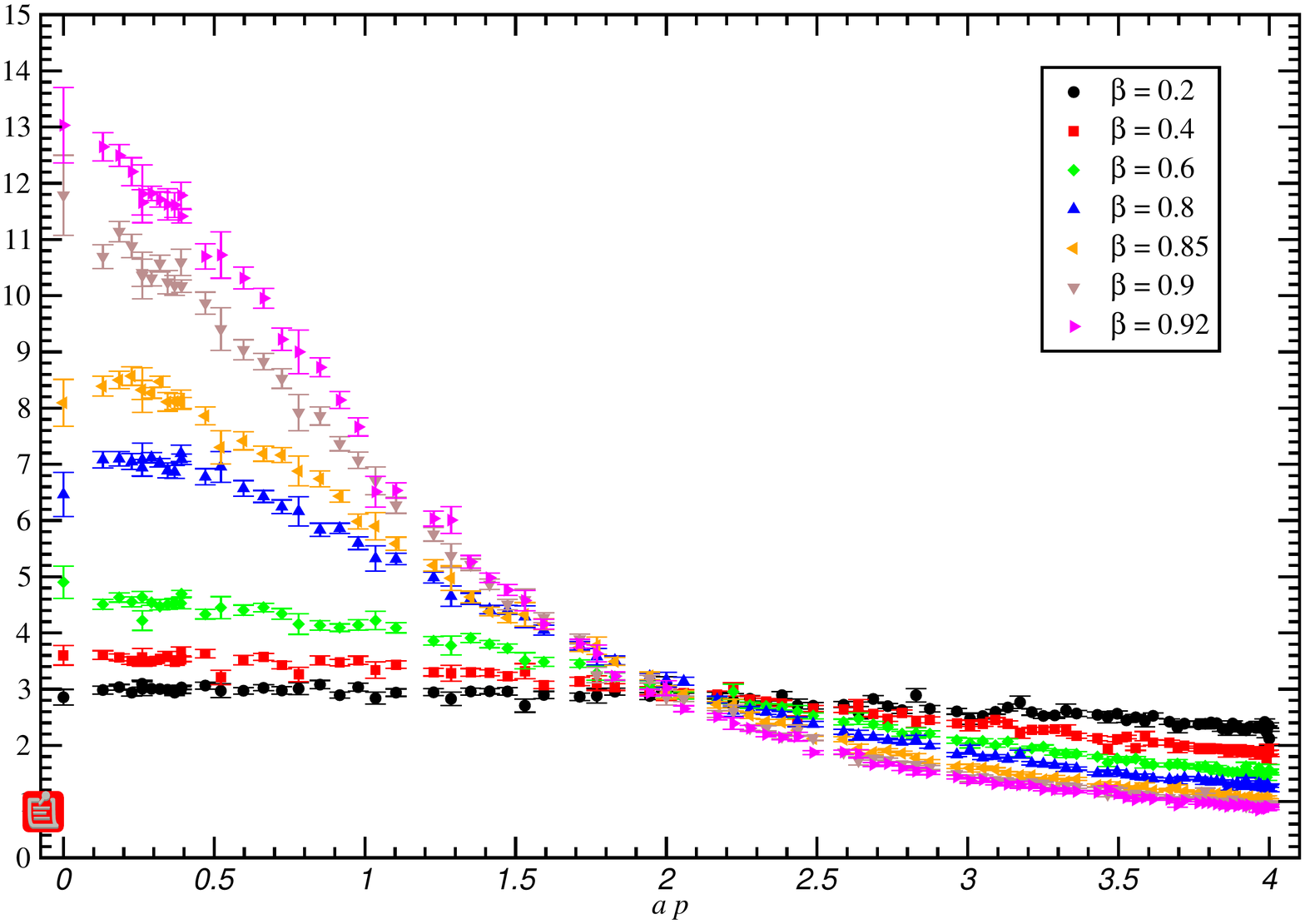}  ~
   \includegraphics[width=3.2in]{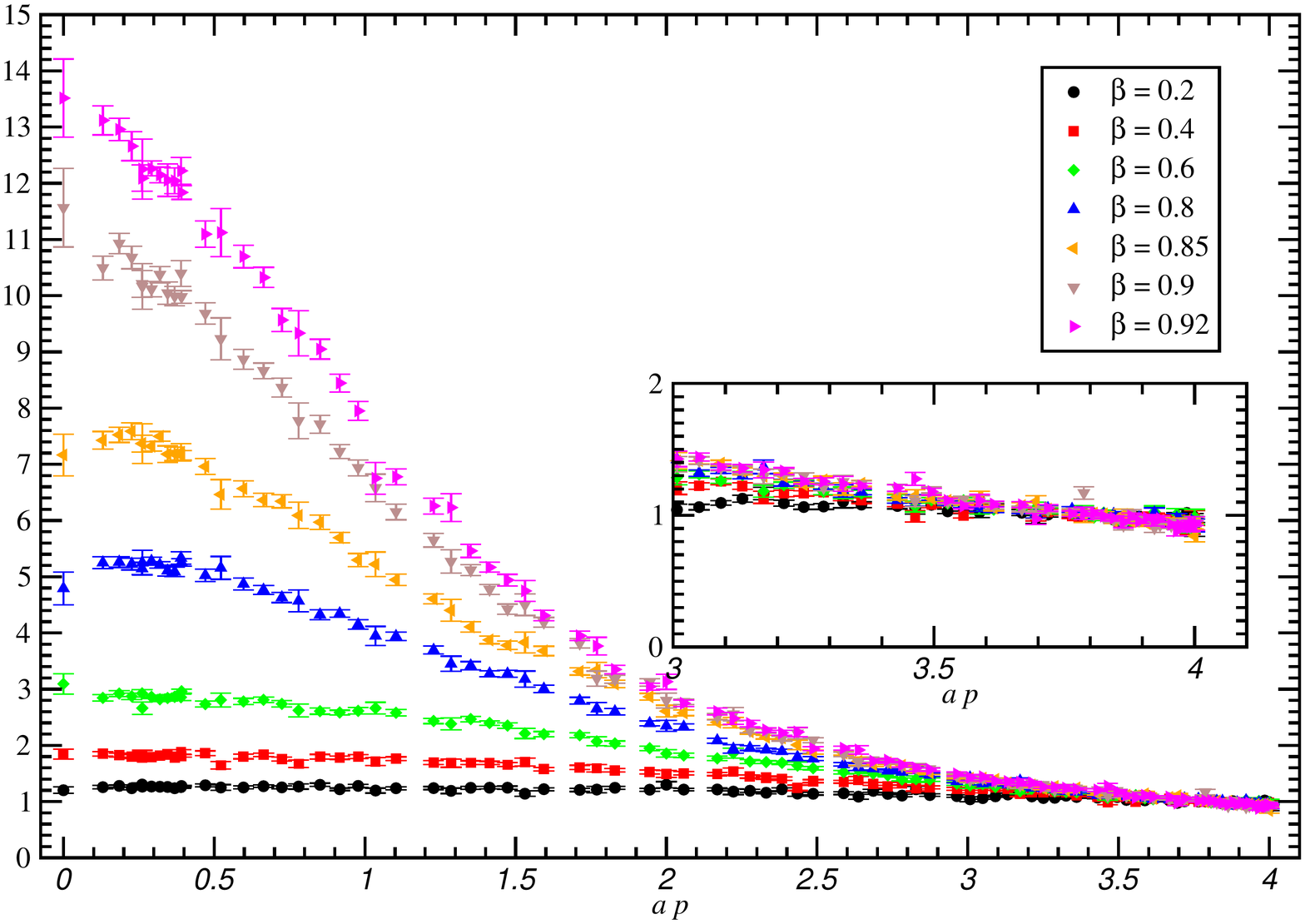}  \\
   \includegraphics[width=3.2in]{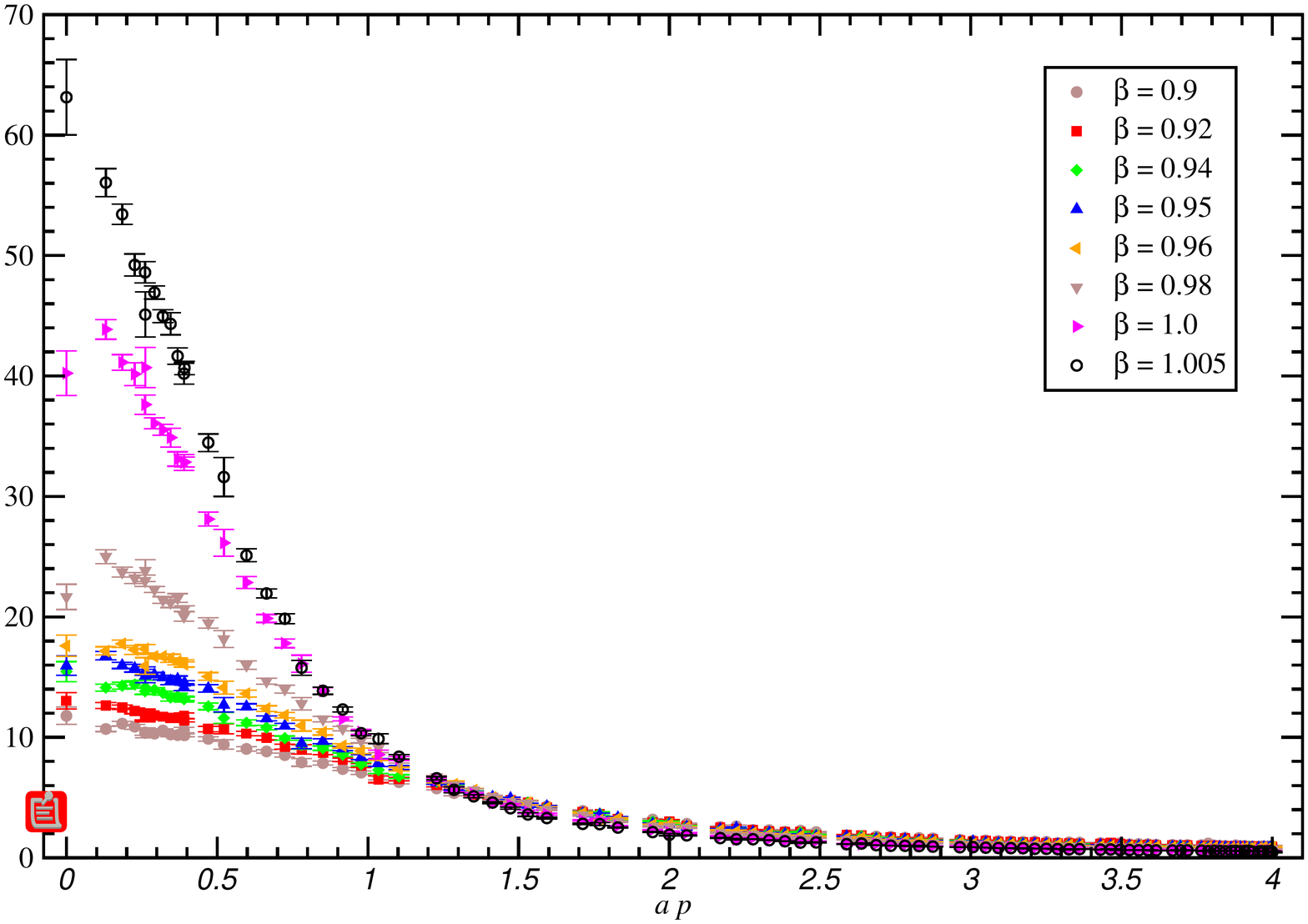}  ~
   \includegraphics[width=3.2in]{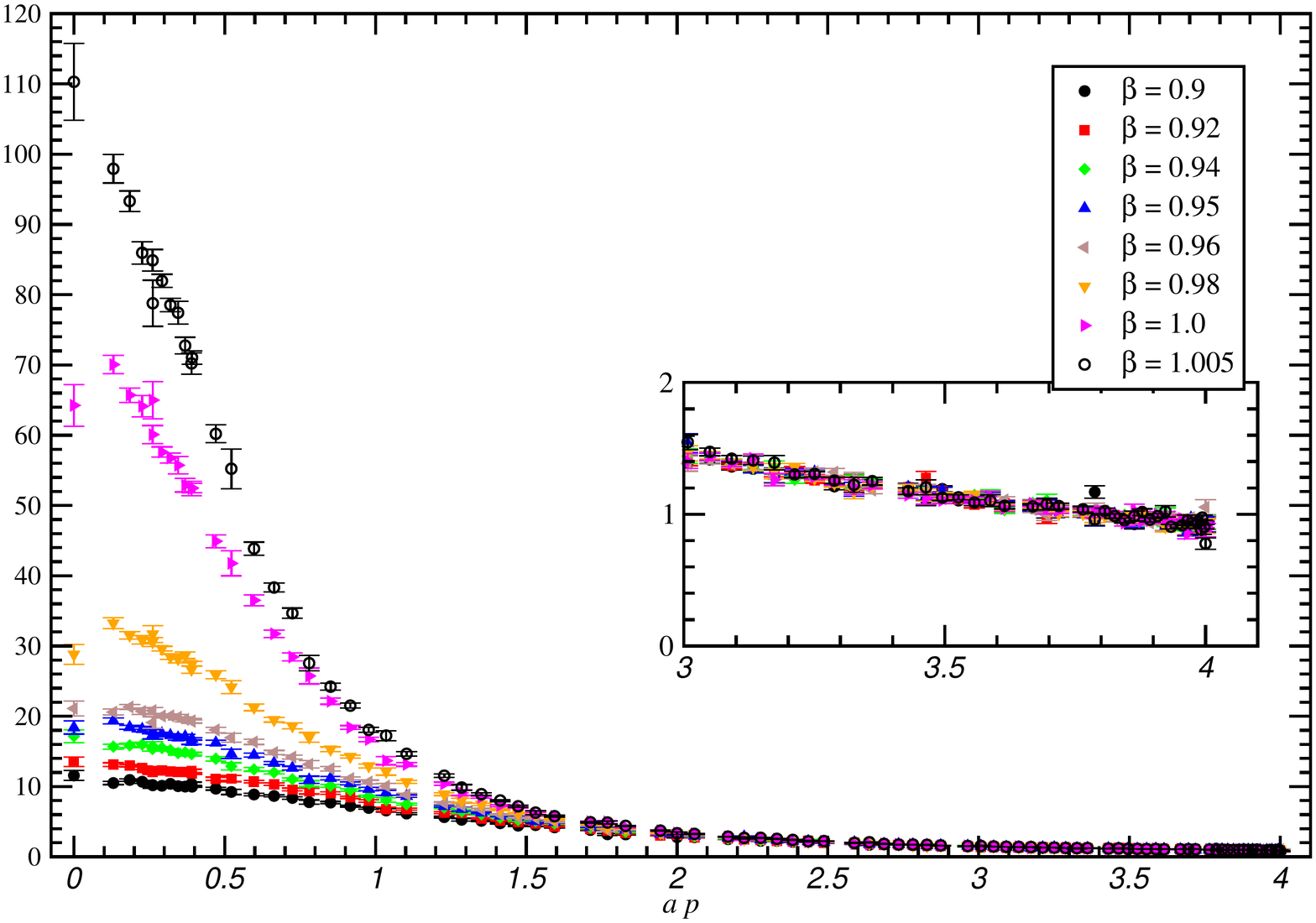}  \\  
   \includegraphics[width=3.2in]{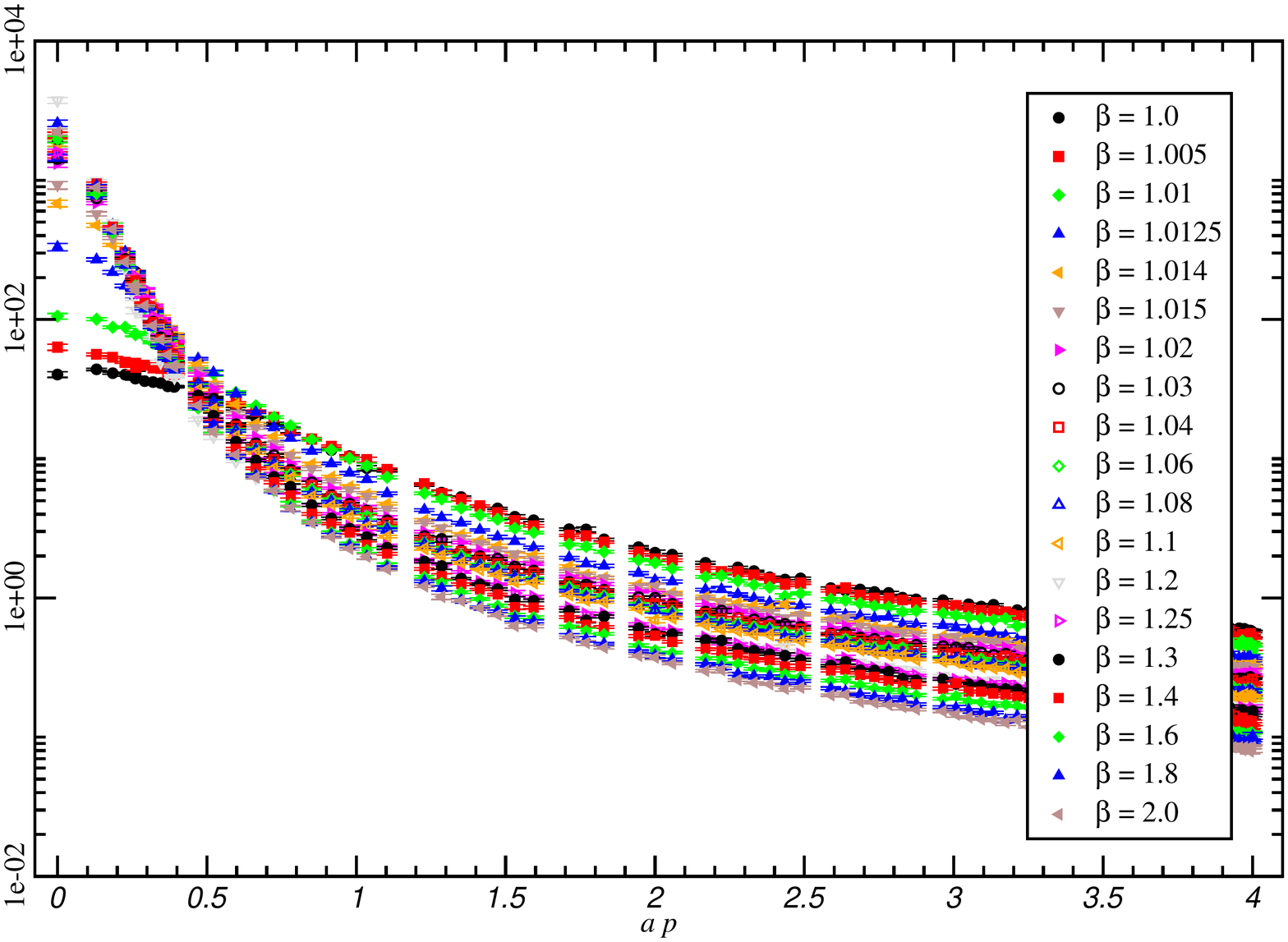} ~ 
   \includegraphics[width=3.2in]{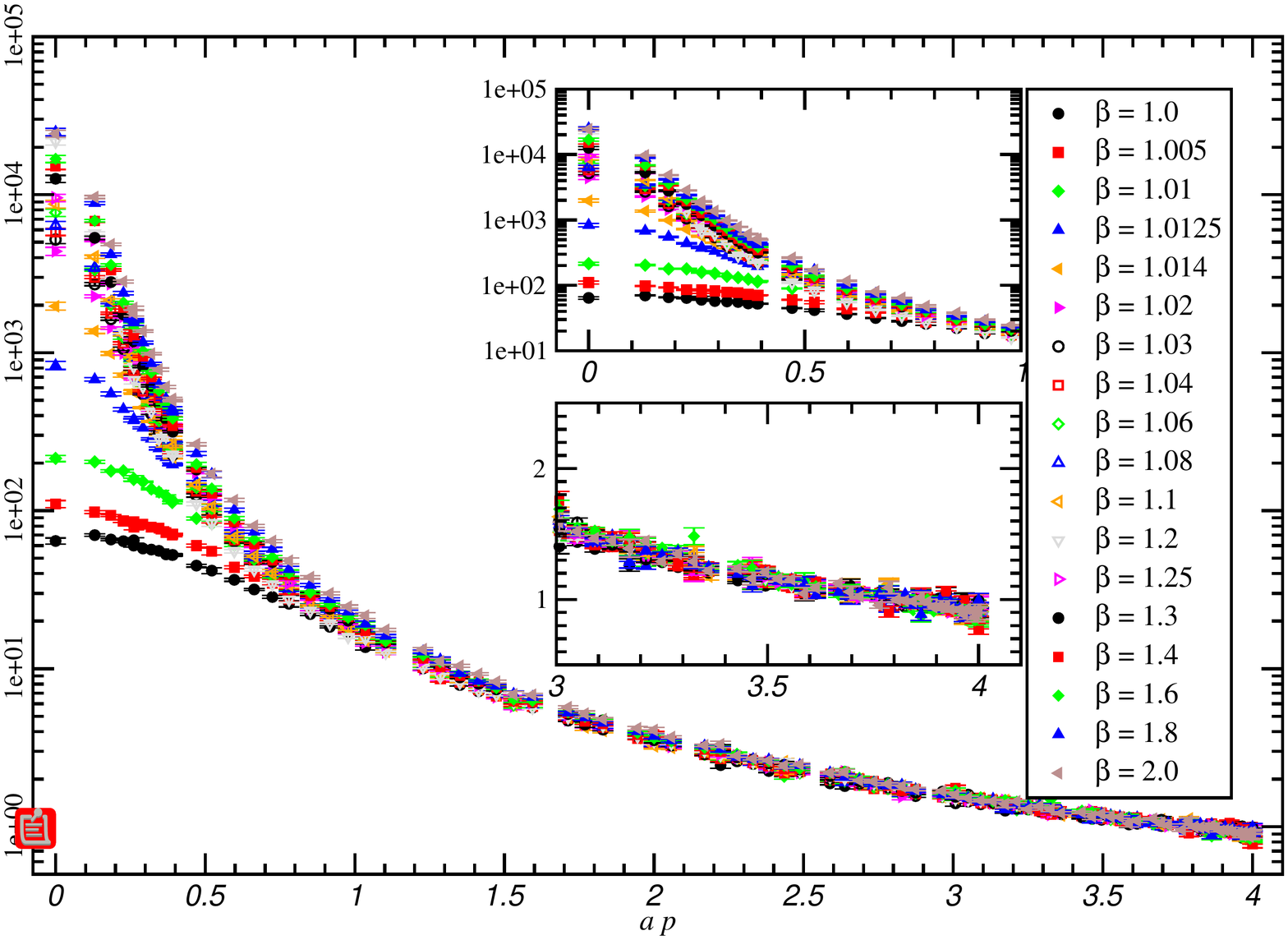}    
   \caption{Dimensionless photon propagator as a function of $\beta$ for the momenta selected by the cuts. Note the differences in the vertical scale.
                 In the bottom plots the vertical scale is logarithmic. The left graphs report the bare photon propagator, while the right graphs show the renormalized
                 photon propagator as described in the main text. For the renormalized propagator the inserted plots include only the high momenta, with the
                 exception of the bottom one that also shows the low momenta propagator for the larger $\beta$ values considered. }
   \label{fig:bare_photon_allBeta}
\end{figure*}

\section{The photon propagator for various $\beta$ \label{SecProp}}

\begin{figure}[t] 
   \centering
   \includegraphics[width=3.1in]{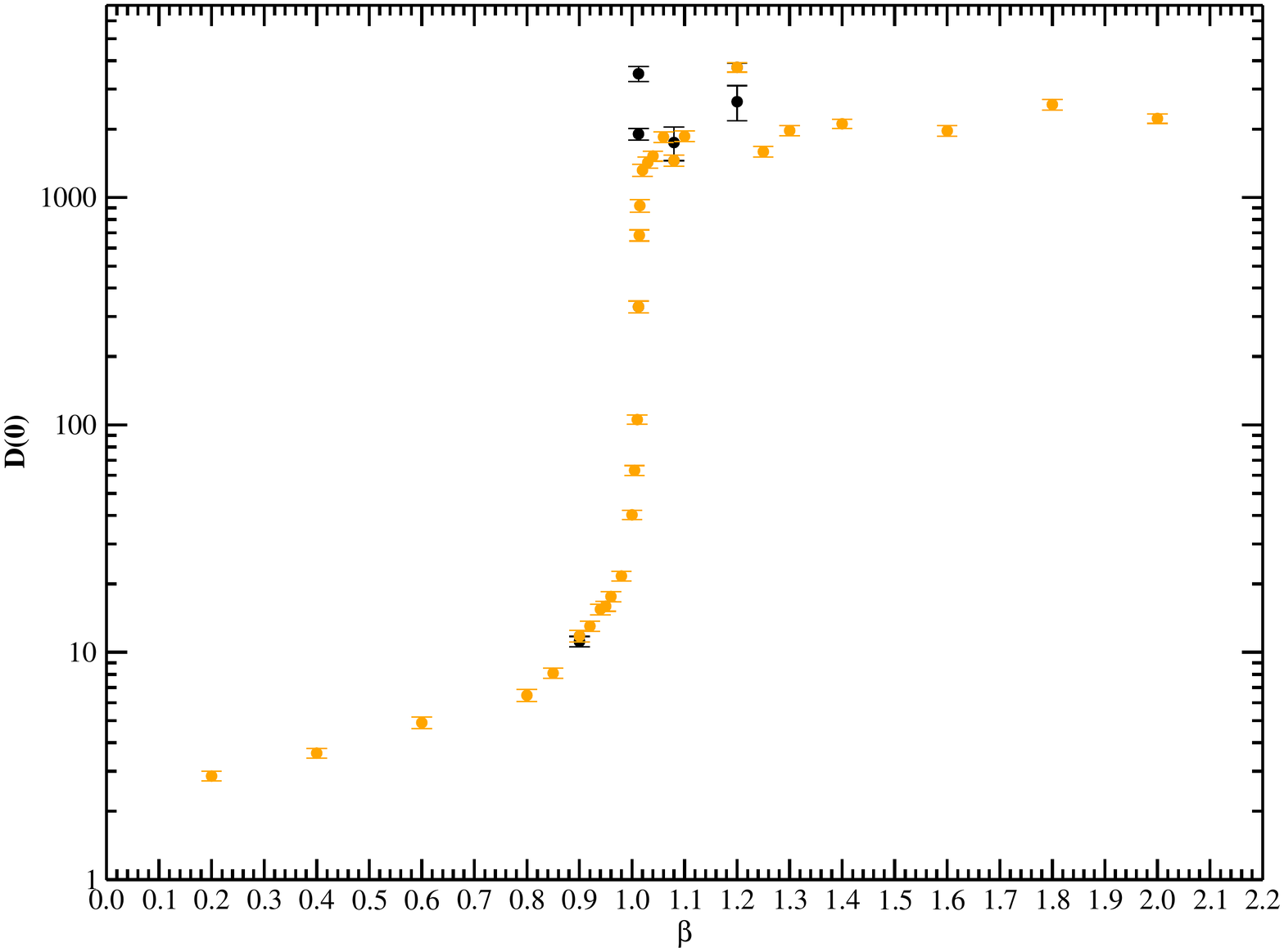}  \\
   \includegraphics[width=3.1in]{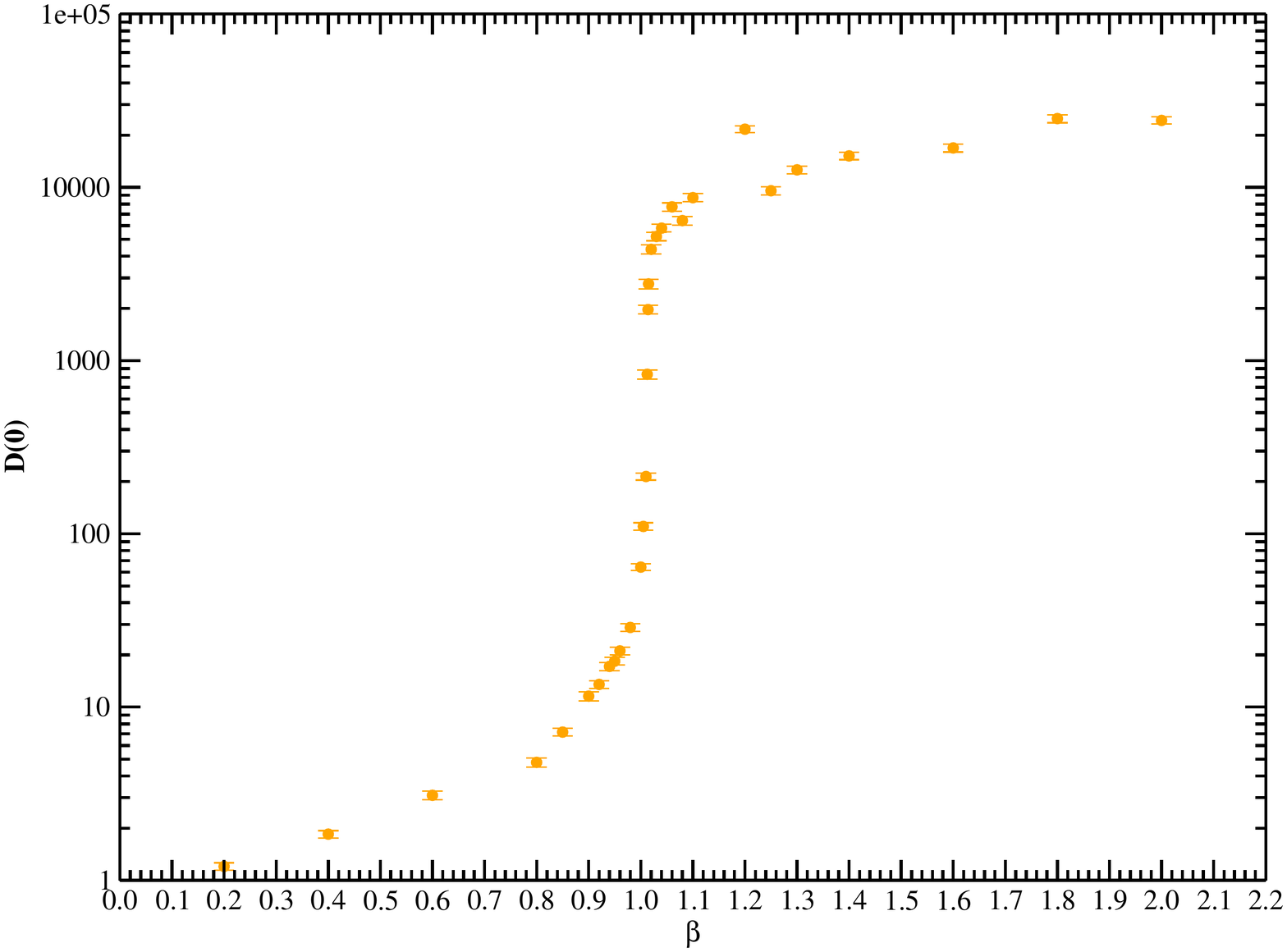}
   \caption{Dimensionless zero momentum bare  (upper plot) and renormalized (lower plot)
                photon propagator for various $\beta$. Note the logarithmic scale for the vertical axis. The points associated with the same $\beta$ 
                but different color are the outcome of different sampling histories (see text for further details).}
   \label{fig:D0_beta}
\end{figure}

In this work we measure the photon propagator for $\beta$ ranging from 0.2 up to 2.0, covering in detail the region $\beta \approx 1$ where the
transition between the two phases is expected to occur.

 The bare photon propagator form factor $D(\hat{p})$ is reported in 
Fig. \ref{fig:bare_photon_allBeta} as a function of the dimensionless improved momentum. The various plots have different vertical scales
and, in particular, for the higher $\beta$  the vertical scale is logarithmic. 
In Fig. \ref{fig:D0_beta} (upper plot) the bare $D(0)$ is shown as a function of $\beta$. 
For the deconfined phase the plots also include the results of different simulations that where performed starting the Markov chain differently. 
The data just referenced is bare lattice data and, therefore, its comparison should be done with care. 
The bottom plot in Fig. \ref{fig:D0_beta} refers to renormalized propagators; more on renormalization later.
This bottom plot reports only the data associated with the orange points in the upper (unrenormalized) plot. 

As  Fig. \ref{fig:D0_beta}  shows, for large $\beta$ there is some dependence of $D(0)$ on the starting configuration used to start the Markov chain;
see the points marked with different colors for the the same $\beta$. However,
the overall result, i.e. the sharp transition from a low $D(0)$ to a large $D(0)$, is not spoiled by the problems that can be ascribed to the sampling method.  

The data in Fig. \ref{fig:bare_photon_allBeta} includes only the results of the runs that are reported in Fig. \ref{fig:D0_beta} in orange.
The bare data in Fig. \ref{fig:bare_photon_allBeta} suggests that the photon propagator is enhanced at low momenta, when
compared to the higher momenta,  for all $\beta$ values, with the enhancement of $D(0)$ increasing with $\beta$.
Moreover, comparing the various plots there seems to be a change in the functional form of $D(\hat{p})$ at $\beta \approx 1.0125$, with
$D(\hat{p})$ becoming much steeper in the low momentum region for larger $\beta$. For $\beta \gtrsim 1.0125$ the propagator
becomes steeper and steeper as $\beta$ increases. This behaviour hints that above this $\beta$ value, the photon propagator is divergent
in the thermodynamical limit and, hopefully, recovers the $1/p^2$ perturbative behaviour of a free field theory. 
The analysis of the volume dependence of the $\beta =1.2$ 
Landau gauge photon propagator data of \cite{Loveridge:2021qzs} supports this claim.

The bare lattice data for $D(0)$ in Fig. \ref{fig:D0_beta} also shows a sharp variation when $\beta$ goes from $\sim 0.8$ 
to $\sim 1.1$ that is associated with the transition to the deconfined continuum-like phase. 
It is possible to identify two baselines for $D(0)$ in this Fig. for $\beta \lesssim 0.8$ and for $\beta  \gtrsim 1.1$. Note however, that for
large $\beta$ the curve associated with $D(0)$ does not seem to be so well defined due to problems with the sampling method.
Despite these problems the presence of a much higher typical value for $D(0)$ for $\beta > \beta_c$ survives to the tests that were performed.

In order to compare the propagators for the different $\beta$ values, one should renormalize the lattice data. The problem for pure gauge
QED being that
there is no clear way to set the scale for its lattice formulation, which makes it impossible to choose a given momentum scale to renormalize
the photon field and, therefore, one has to look for alternatives. As a tentative method to overcome the problem of the scale setting, 
we choose to fix the high momentum behaviour of the photon propagator to renormalize the theory assuming that $D(\hat{p})$ is independent of
$\beta$ in the ultraviolet region. In order to use such a procedure one has to assume that in all the simulations
the asymptotic ultraviolet region is accessed for all $\beta$ values. As the results show, see the insertions in Fig. 
\ref{fig:bare_photon_allBeta}, this assumption seems to be validated by the lattice data.

As a first step towards renormalizing the lattice data, for each $\beta$ value the data for momenta in the range $a \, \hat{p} \in [ 3.5 \, , \, 4]$ 
is fitted to the following functional form
\begin{equation}
   D_{\text{Fit}} (x)  = a + \frac{b}{x^2} \ .
\end{equation}   
Typical values for the $\chi^2/d.o.f.$ for the fits are below 1.6, with 0.62 and 2.2 being the smallest and highest observed figures, respectively. 
Then, for each $\beta$, the corresponding fit is used to set $D(a \mu) = 1$ at $a \mu = 3.8$. Recall that $a \hat{p} = 4$ is the highest dimensionless 
momentum accessed in the simulation and our choice of $a \mu$ is the middle point of the fitting range.

Similarly as observed for the bare lattice data, the renormalized propagators show  a drastic change in the low momentum region for
$\beta  \gtrsim 1.0125$.
The data in Fig. \ref{fig:D0_beta} shows a transition from low $\beta$ and low $D(0)$, that starts starts at $\beta \approx 0.8$ and
ends at $\beta \approx 1.1$, with $D(0)$ taking relatively large values for $\beta  \gtrsim  1.1$ (note the log scale on the vertical axis).
Note also that again for $\beta  \approx 1.1$ and above there seems to be some dispersion on the values of $D(0)$ that are, in principle,
due to the use of the hybrid Monte Carlo method as sampling method.

The photon propagator data shows that this two point correlation function, through the form factor $D(\hat{p})$,
can be used to distinguish the phases of pure gauge compact QED. It follows that the dynamics of the gauge fields in the confined
and deconfined phases are rather different. 

Our conclusion for the photon propagator is inline with similar studies for QCD where the gluon propagator was studied as a function 
of the temperature \cite{Aouane:2011fv,Silva:2013maa}. 
Indeed, these studies shows that the gluon propagator can be used to distinguish the confined and deconfined phases
and some authors suggested that $D(0)$ can be used as an order parameter for the deconfinement transition.
Furthermore, for the deconfined region the gluon propagator can also be used to distinguish
the topological sectors that are associated with  the center symmetry in the pure gauge Yang-Mills theory \cite{Silva:2016onh}.

\section{Dirac strings and the Confinement and Deconfinement Phases \label{Sec:DiracStrings}}

\begin{figure}[t] 
   \centering
   \includegraphics[width=3.1in]{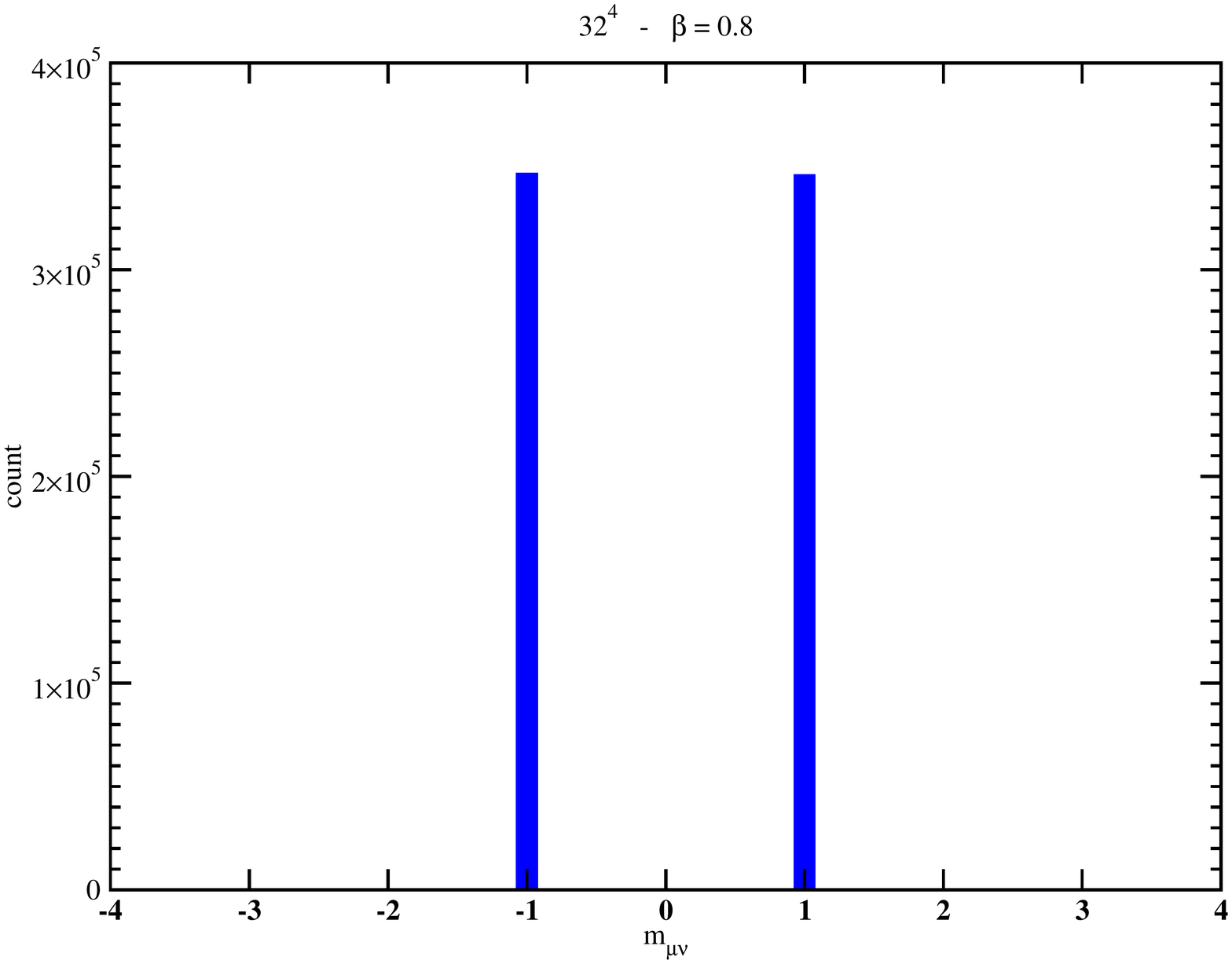}  \\
   \includegraphics[width=3.1in]{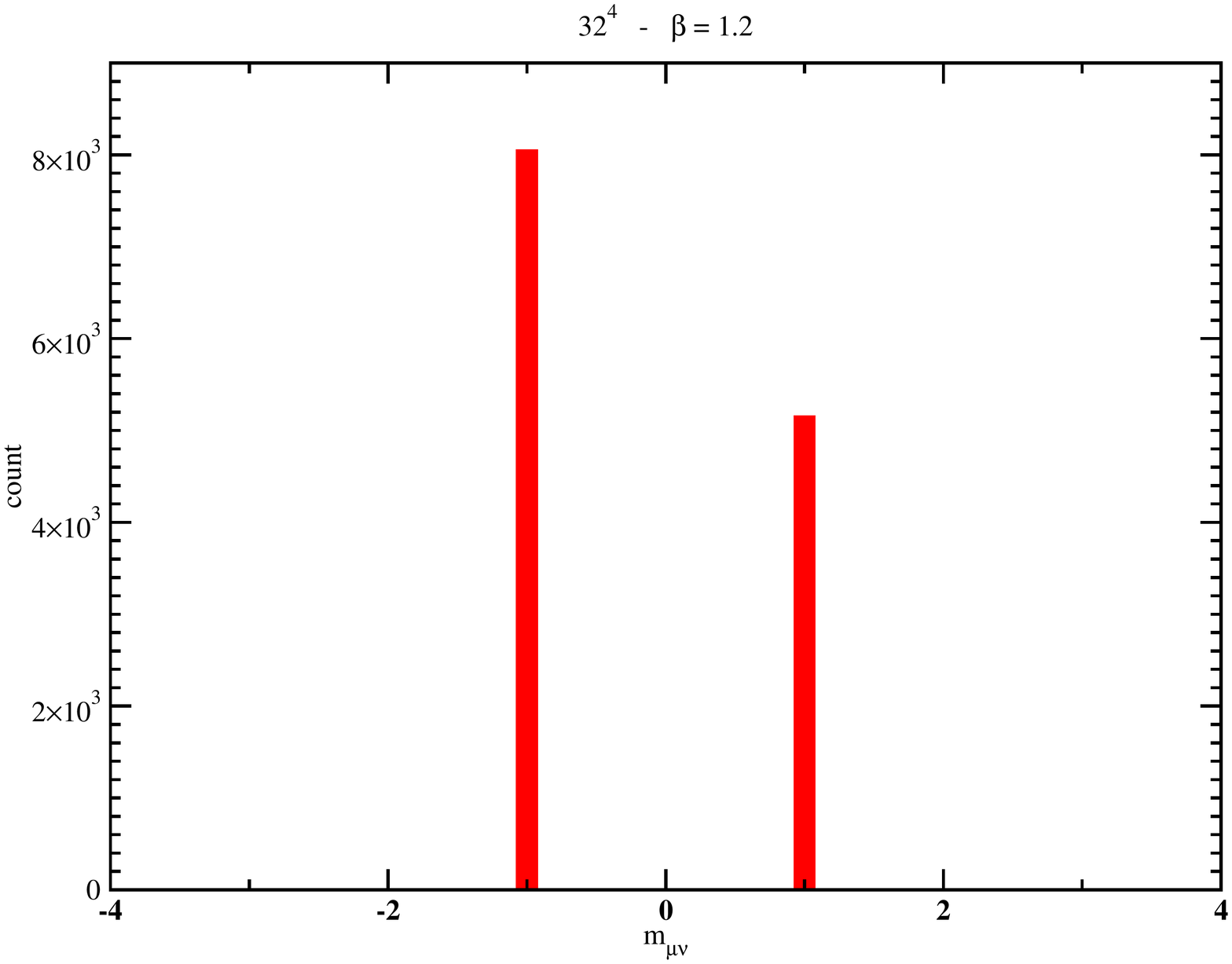}
   \caption{Dirac string distribution on a $32^4$ lattice for the confined (upper plot) and deconfined (bottom plot) phases. The value
   $m_{\mu\nu} (x) = 0$  was removed from the histogram as its contribution is much larger than those reported in the plots. See text for details.}
   \label{fig:monopols_beta}
\end{figure}

\begin{figure}[t] 
   \centering
   \includegraphics[width=3.1in]{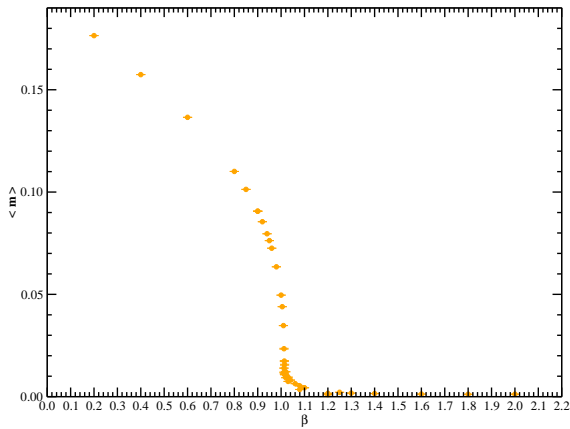} 
   \caption{Ensemble average of the mean density of Dirac strings as a function of $\beta$.}
   \label{fig:m_abs_beta}
\end{figure}

The integer field $m_{\mu\nu} (x)$ is related to the topology of the gauge fields as it measures the presence of Dirac strings in the gauge configurations. 
For the two classes of gauge configurations considered in \cite{Loveridge:2021qzs}, those in the confined phase ($\beta = 0.8$) and
those in the deconfined phase ($\beta = 1.2$), $m_{\mu\nu} (x)$ was measured  and in 
Fig. 9 of \cite{Loveridge:2021qzs} the average value over the lattice of $|m_{\mu\nu}|$ for the Landau gauge configurations for $\beta = 0.8$ and 
$\beta = 1.2$ is illustrated. 
For these $\beta$'s the average value over the lattice of $|m_{\mu\nu}|$ 
differs by two orders of magnitude, with the larger value being associated with the configurations in the confined phase ($\beta = 0.8$). 
This was taken as an indication that Dirac strings play a major role in the dynamics of the confined phase and are at the origin of the observed mass gap.

In Fig. \ref{fig:monopols_beta}  we report the typical distribution of the integer field $m_{\mu\nu}$ for $32^4$ gauge configurations
in either of the phases. The upper plot refers to a configuration in the confined phase, while the lower plot is for
a configuration in the deconfined phase. In both cases the averages over the lattice of $m_{\mu\nu}$ are compatible with zero.
In Fig. \ref{fig:monopols_beta} we do not show the bar associated with $m_{\mu\nu} = 0$ that clearly dominates the distribution.
Indeed, for the configuration in the confined ($\beta = 0.8$) phase the number of vanishing $m_{\mu\nu}$ is about 16 times the number of
$m_{\mu\nu} = \pm 1$, while for deconfined phase this ratio increases to about 695. 
In the deconfined phase, the number of vanishing $m$ is larger by a factor of $\sim 43$ relative to the confined case
and the distribution of $m \ne 0$ becomes asymmetric. 
These observations explains the pattern of the non-gauge invariant quantity 
\begin{equation}
m = \frac{1}{6V} \sum_{x, \, \mu < \nu} \, | m_{\mu\nu} (x) |
 \end{equation}
 reported in \cite{Loveridge:2021qzs}, see their Fig. 9,
and why $m$ decreases drastically in the deconfined phase in comparison with the confined phase.

Our new simulations offer an opportunity to monitor how the number of Dirac strings evolves with $\beta$ and its connection with the generation of the 
photon mass gap. We remind the reader that the mass gap is identified with a finite $D(0)$ but no attempt is made to measure its value.

Fig \ref{fig:m_abs_beta} reports on the ensemble averages of $m$, named $\langle m \rangle$,  as a function of $\beta$. This quantity
takes larger values in the confined phase and smaller $\beta$ values, it has a sudden drop around $\beta \sim 1$ and fluctuates at rather small values for 
$\beta \gtrsim 1$. To illustrate its quantitative behaviour,  the quantity  $\langle m \rangle$ goes from $0.1764749(47)$ at $\beta = 0.2$ to $1.206(18) \times 10^{-3}$ for 
$\beta = 1.2$ that represents a reduction by a factor of $\sim 146$. 

The comparison of Figs.  \ref{fig:m_abs_beta} and \ref{fig:D0_beta} establishes a correlation between $D(0)$ and $\langle m \rangle$.
Indeed, $D(0)$ follows the opposite behaviour of $\langle m \rangle$ with a fast change and an almost infinite slope, taking place at the 
same $\beta$ values. 
The sudden increase of $D(0)$, that can be translated into a sudden decrease of the photon mass gap,
occurs exactly when $\langle m \rangle$ drops and becomes close to zero. 
This behaviour suggests, once more, that the Dirac strings have a primordial role 
in the confined phase. The correlation between the evolution with $\beta$ of  $D(0)$ and $\langle m \rangle$ suggests that the Dirac strings are also at the 
origin of the photon mass that is non-vanishing only in the confined phase.

\subsection{The hybrid Monte Carlo sampling and the photon propagator \label{Sec:HMC-DiracStrings}}

As mentioned  previously, the hybrid Monte Carlo method seems to have problems in sampling the Wilson action for compact QED. 
To check for possible bias, for a number of $\beta$s several simulations were run using different gauge configurations to start
the Markov chain, and then compare the corresponding propagators and the associated topological structures as measured by $\langle m \rangle$.

\begin{figure*}[t] 
   \centering
   \includegraphics[width=3.4in]{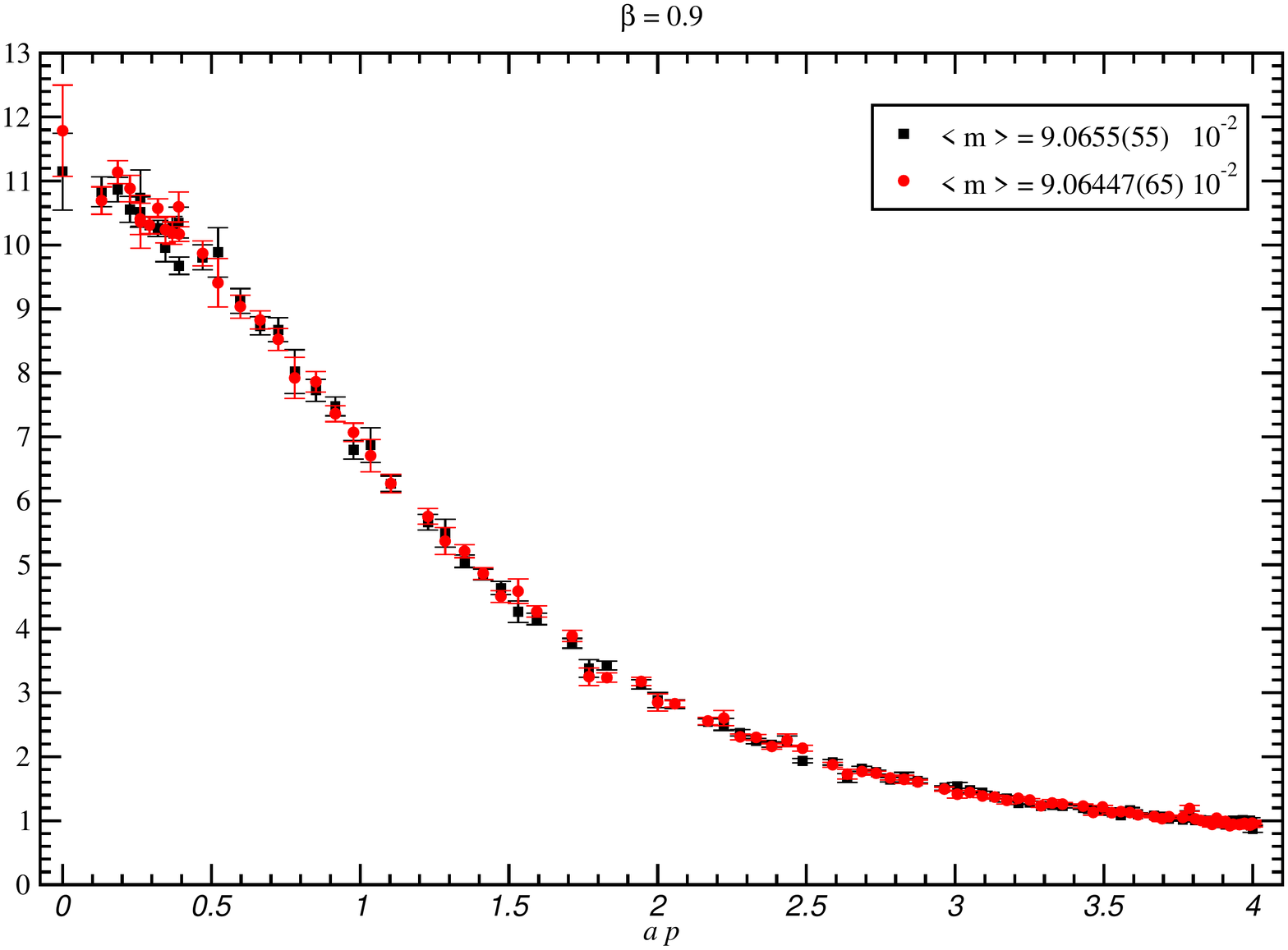}  ~ ~ 
   \includegraphics[width=3.4in]{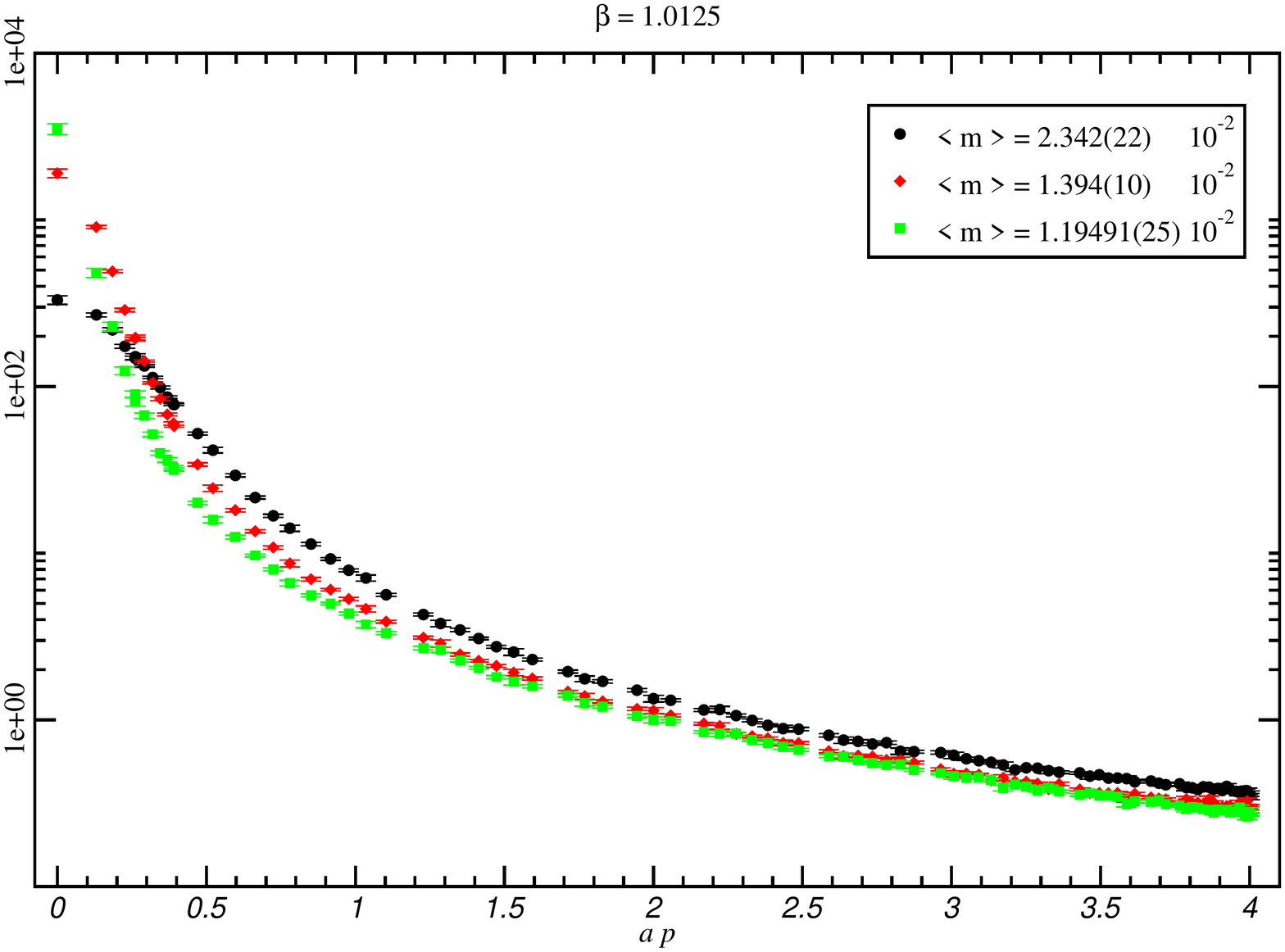} \\
   \includegraphics[width=3.4in]{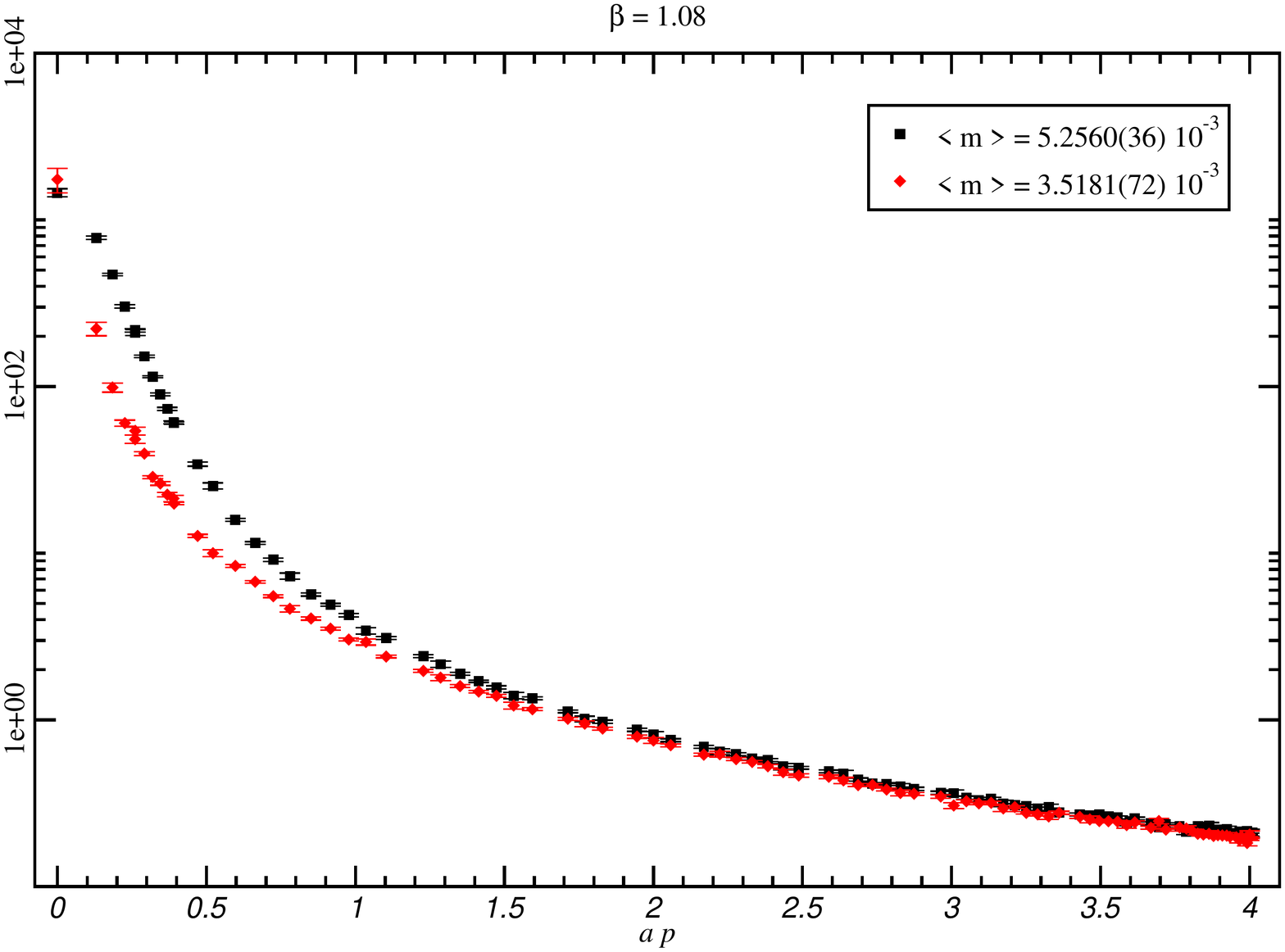}  ~  ~
   \includegraphics[width=3.4in]{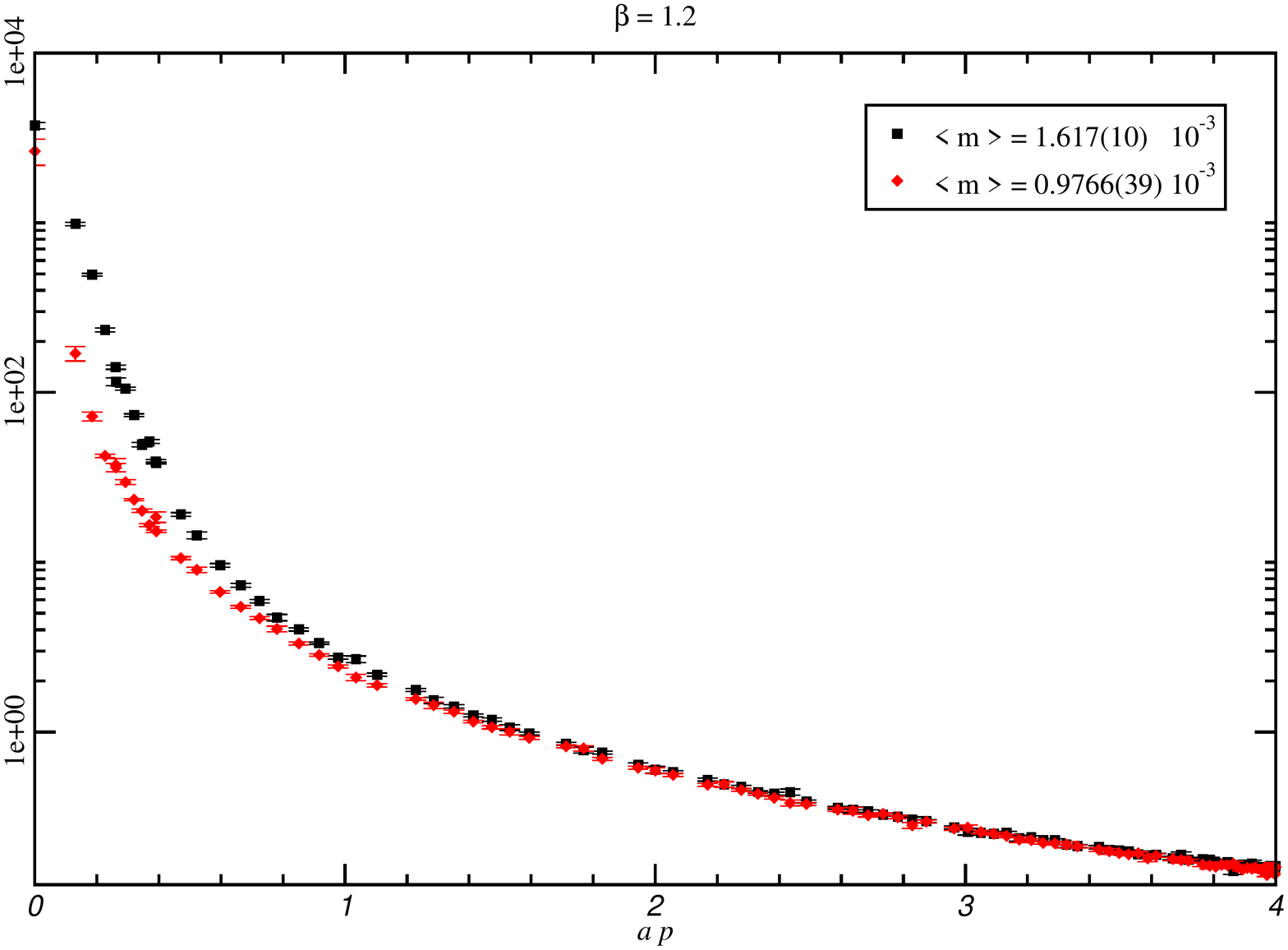} 
   \caption{Bare  dimensionless data for $D(p^2)$ and for several $\beta$ from simulations using a different starting point in the 
   Markov chain. Note the logarithmic scale for the plots associated with $\beta > 1$.}
   \label{fig:photon_prop_and_hmc}
\end{figure*}

\begin{figure*}[t] 
   \centering
   \includegraphics[width=3.4in]{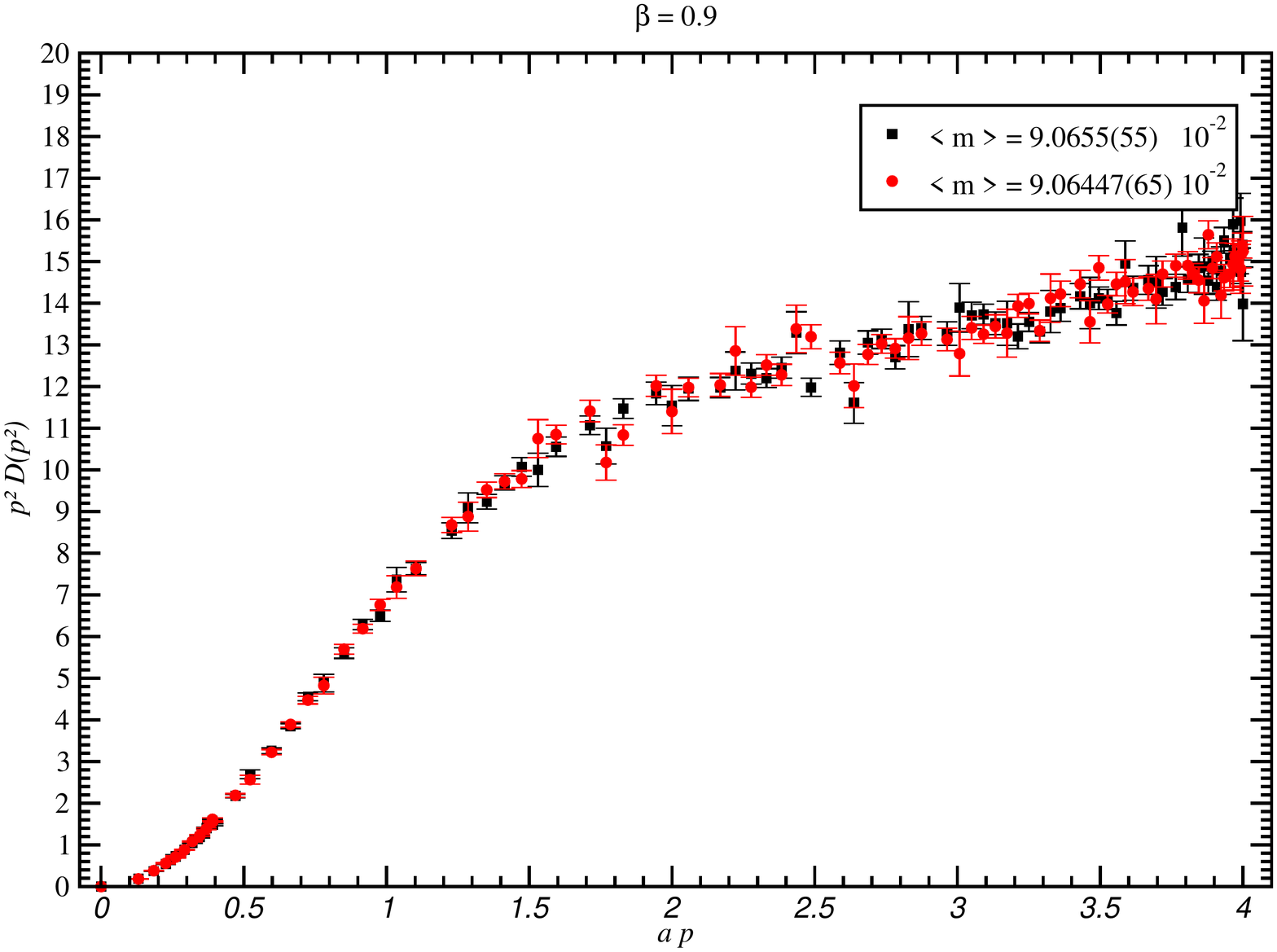}  ~ ~ 
   \includegraphics[width=3.4in]{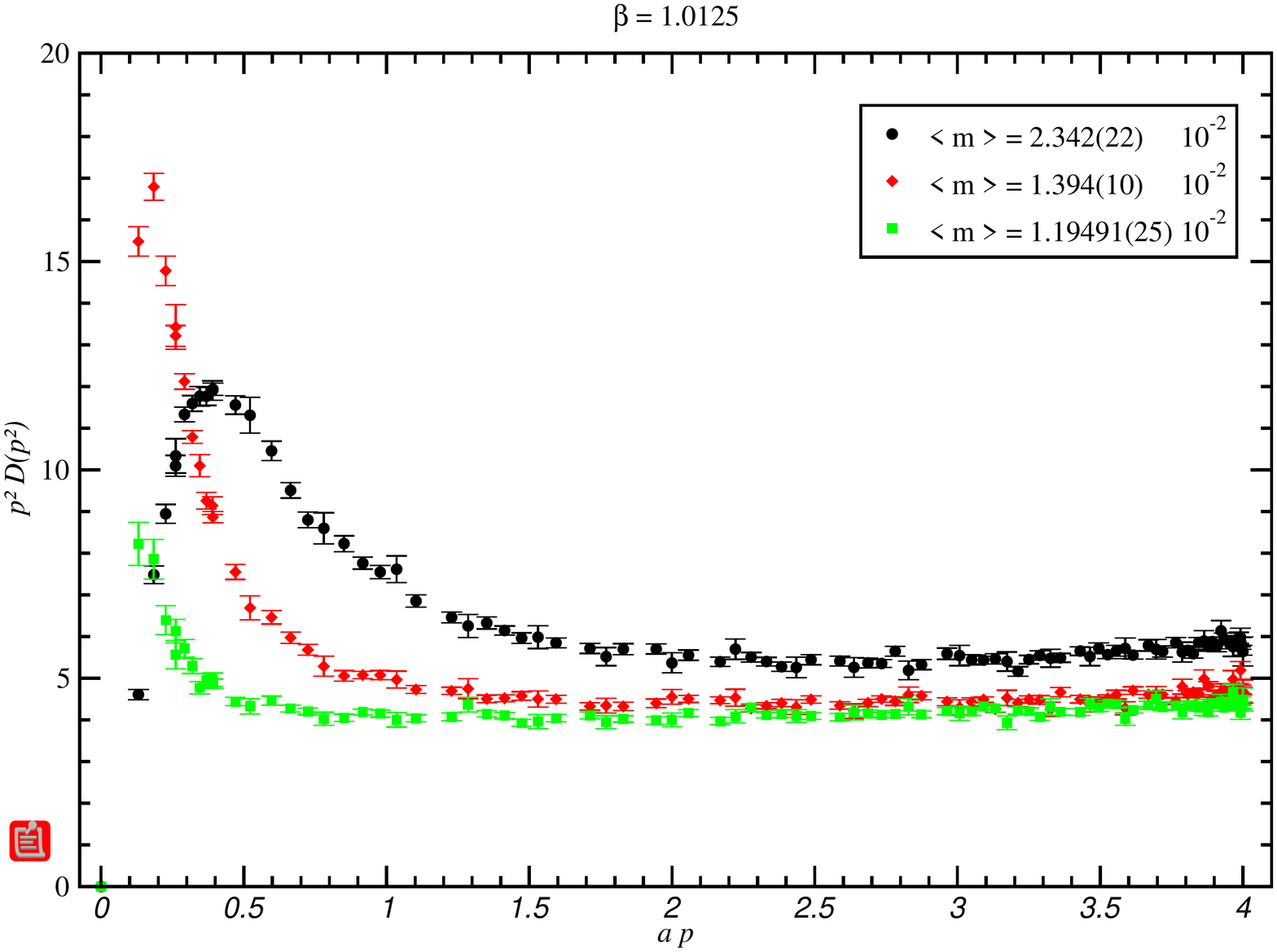} \\
   \includegraphics[width=3.4in]{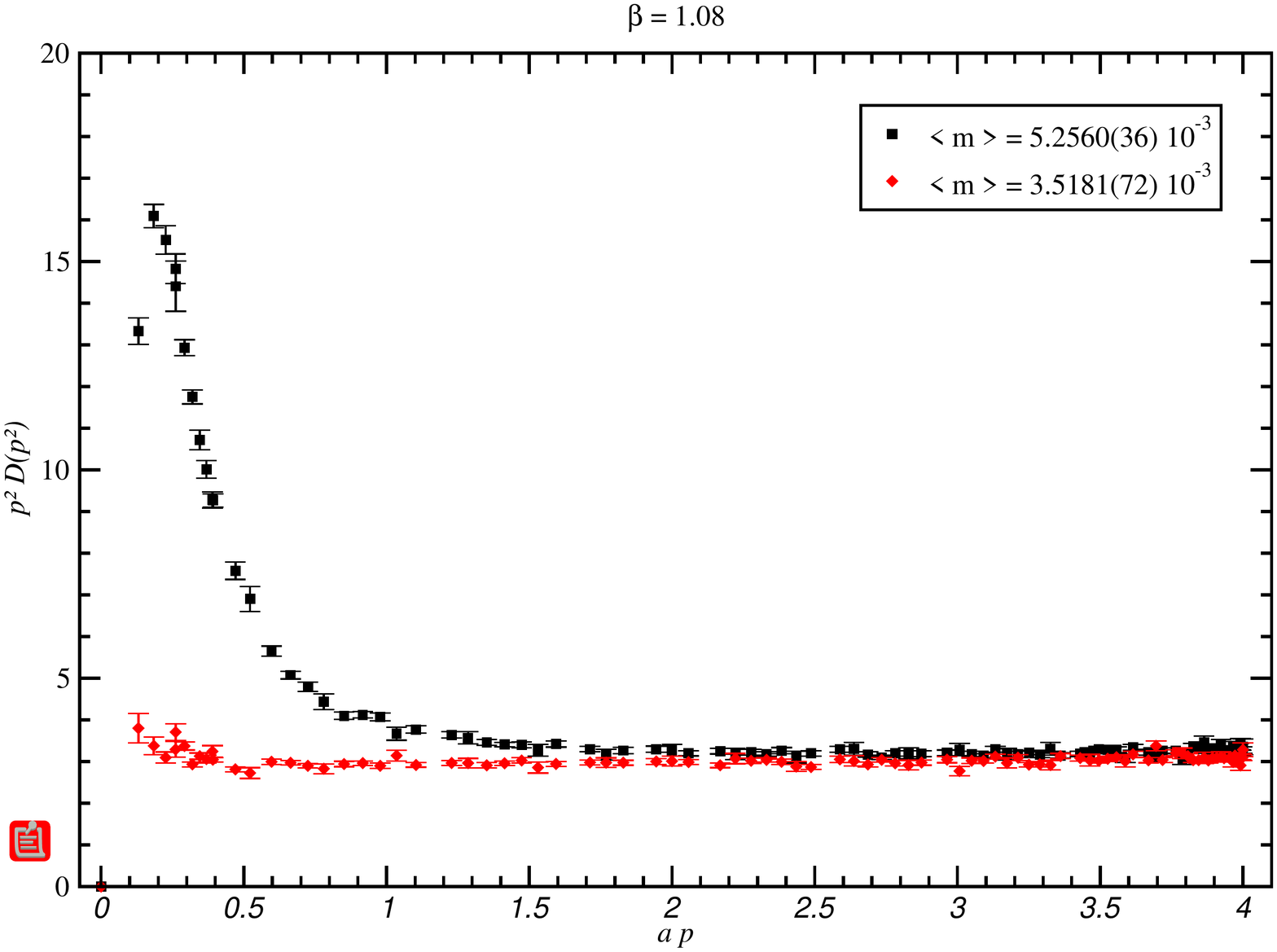}  ~  ~
   \includegraphics[width=3.4in]{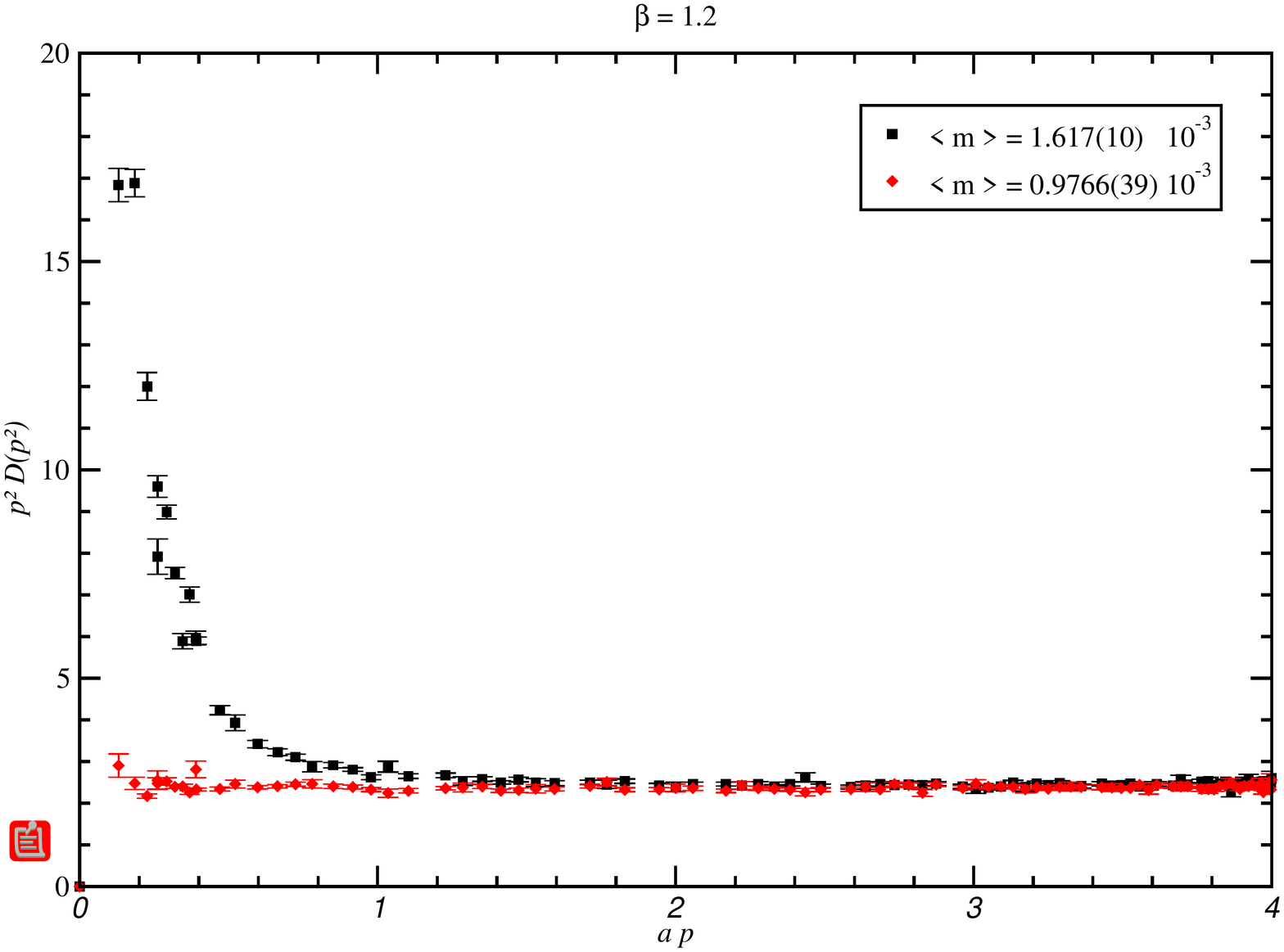} 
   \caption{Bare dimensionless data for $p^2 D(p^2)$ and for several $\beta$ from simulations using a different starting point in the Markov 
   chain. }
   \label{fig:photon_prop_p2_and_hmc}
\end{figure*}

In Fig. \ref{fig:photon_prop_and_hmc} we report on the  photon propagator computed at several $\beta$ using the results of
the Monte Carlo simulations initiated with different starting points. 
For the smallest $\beta$ value, that is below the confinement-deconfinement transition, the photon propagators coming from the two simulations are equal 
within one standard deviation, both evaluations of the propagator are finite at zero momentum and the corresponding $\langle m \rangle$'s are  compatible
within  errors. This suggests that below the confinement-deconfinement transition the hybrid Monte Carlo method is robust and does a proper 
sampling of the compact QED action.

On the other hand, the simulations performed in the deconfined phase, i.e. for $\beta > 1$, show that  the evaluation of the photon propagator with
the hybrid Monte Carlo method is 
(i) sensitive to the initialization of the Markov chain, although in all cases the propagator increases and seems to diverge as one approaches 
the $p \rightarrow 0^+$ limit, 
and (ii) for the ensembles with the same $\beta$ those  with smaller average number of Dirac strings seem to have a propagator 
that is closer to the free field propagator $1/p^2$ behaviour in the infrared region. 
This is better seen  in Fig. \ref{fig:photon_prop_p2_and_hmc} where the photon dressing function $p^2 D(p^2)$ is plotted. For a free field type of propagator
the dressing function should be constant.

As Figs. \ref{fig:photon_prop_and_hmc}  and \ref{fig:photon_prop_p2_and_hmc} show, in the deconfined phase
the configurations with the smallest $\langle m \rangle$ have a dressing function
that is essentially flat and it is in this sense that the propagator is closer to a free field theory propagator. The deviations from a $1/p^2$ propagator
are larger in the infrared region. Furthermore, these Figs. also show that it is closer to the confinement-deconfinement transition that
the hybrid Monte Carlo performs worst, as this is where the propagators differ the most. 
See, in particular, the propagators computed from the various simulations at $\beta = 1.0125$.

\section{Summary and Conclusions \label{Sec:Summary}}

In this work the lattice Landau gauge photon propagator together with the average number of Dirac strings
is studied in the compact formulation of QED for the pure gauge version of the theory. 
Following the procedure implemented in  \cite{Loveridge:2021qzs}, the confined phase and deconfined phase propagators are computed. 
Our results show a correlation between the two quantities that clearly identify the confinement-deconfinement transition. 
The analysis of the propagator data suggests that it is for $ \beta \ge 1.0125$ that photon propagator becomes closer to a free  propagator.

The simulations also show that the functional form of the photon propagator is different in each phase. 
In the confined phase the propagator is finite over the full range of momenta, an indication that the theory has a mass gap for low $\beta$.
For the deconfined phase the photon propagator becomes compatible with a divergent propagator in the infrared region, 
suggesting that it reproduces a free field like propagator in the thermodynamic limit. 
Note that in the present work we do not investigate the thermodynamic limit of the theory but this 
conclusion comes from combining the new results with those of \cite{Loveridge:2021qzs}.

The differences observed in the photon propagator as a function of $\beta$ correlated well with the average number of Dirac strings present
in the gauge configurations, i.e. with the topological structure of the gauge group. Indeed, the simulations show that the presence of a large number 
of Dirac strings results in a theory with a mass gap that vanishes in the deconfined phase. It is in the deconfined phase where there are essentially 
no Dirac strings found in the gauge configurations.

The results discussed here also show  that both the photon propagator and the number of Dirac strings can be used to distinguish the two phases of
the lattice compact QED formulation. Furthermore, from the analysis of the data for these two quantities, it seems that the transition between the confined
and deconfined phase is first order.

\section*{Acknowledgments}

This work was partly supported by the FCT – Funda\c{c}\~ao para a Ci\^encia e a Tecnologia, I.P., under Projects Nos. 
UIDB/04564/2020 and UIDP/04564/2020.
P. J. S. acknowledges financial support from FCT (Portugal) under Contract No. CEECIND/00488/2017.
The authors acknowledge the Laboratory for Advanced Computing at the University of Coimbra (\url{http://www.uc.pt/lca}) 
for providing access to the HPC resource Navigator.


\end{document}